\documentclass[structabstract]{aa}  
\usepackage{epsfig,amsmath}
\usepackage[dvips]{color}
\usepackage{verbatim}
\usepackage{graphicx}
\usepackage{txfonts}
\usepackage{natbib}
\bibpunct{(}{)}{;}{a}{}{,}
\usepackage{hyperref}

\definecolor{Black}{named}{Black}
\definecolor{Red}{named}{Red}
\definecolor{Green}{named}{Green}
\definecolor{Blue}{named}{Blue}

\begin{document}

 \title{No evidence for dust extinction in GRB\,050904 at $z\sim6.3$}

\author{Tayyaba Zafar\inst{1}
 \and Darach J. Watson\inst{1}
  \and Daniele Malesani \inst{1}
   \and Paul M. Vreeswijk  \inst{1}
    \and Johan P.U. Fynbo\inst{1}
     \and Jens Hjorth\inst{1}
       \and Andrew J. Levan\inst{2}
         \and Micha{\l} J. Micha{\l}owski\inst{1,3}} 

\institute{Dark Cosmology Centre, Niels Bohr Institute, University of Copenhagen,
Juliane Maries Vej 30, DK-2100 Copenhagen \O, Denmark
\and Department of Physics, University of Warwick, Coventry, CV4 7AL, United Kingdom
\and Scottish Universities Physics Alliance, Institute for Astronomy, University of Edinburgh, Royal Observatory, Edinburgh, EH9 3HJ, United Kingdom}

\offprints{tayyaba@dark-cosmology.dk}

\date{Received  / Accepted }

\abstract
 {Gamma-ray burst (GRB) afterglows are excellent and sensitive probes of gas and dust in star-forming galaxies at all epochs. It has been posited that dust in the early Universe must be different from dust at lower redshifts. To date two reports in the literature directly support this contention, one of which is based on the spectral shape of the afterglow spectrum of GRB\,050904 at $z=6.295$.}
  {Here we reinvestigate the afterglow of GRB\,050904 to understand cosmic dust at high redshift. We address the claimed evidence for unusual (supernova-origin) dust in its host galaxy by simultaneously examining the X-ray and optical/near-infrared spectrophotometric data of the afterglow.} 
   {We derive the intrinsic spectral energy distribution (SED) of the afterglow at three different epochs,  0.47, 1.25 and 3.4 days after the burst, by re-reducing the \emph{Swift} X-ray data, the 1.25 days FORS2 $z$-Gunn photometric data, the spectroscopic and $z^\prime$-band photometric data at $\sim3$ days from the Subaru telescope, as well as the critical UKIRT $Z$-band photometry at 0.47 days, upon which the claim of dust detection largely relies.}
    {We find no evidence of dust extinction in the SED at any time. We compute flux densities at $\lambda_{\rm{rest}}=1250\,\AA$ directly from the observed counts at all epochs. In the earliest epoch, 0.47 days, where the claim of dust is strongest, the $Z$-band suppression is found to be smaller ($0.3\pm0.2$\,mag) than previously reported and statistically insignificant ($<1.5\sigma$). Furthermore we find that the photometry of this band is unstable and difficult to calibrate.}
    {From the afterglow SED we demonstrate that there is no evidence for dust extinction in GRB\,050904 -- the SED at all times can be reproduced without dust, and at 1.25 days in particular, significant extinction can be excluded, with $A(3000\,\AA) <0.27$\,mag at 95\% confidence using the supernova-type extinction curve. We conclude that there is no evidence of any extinction in the afterglow of GRB\,050904 and that the presence of supernova-origin dust in the host of GRB\,050904 must be viewed skeptically.} 

\keywords{Galaxies: high-redshift - ISM: dust, extinction - Gamma rays: bursts}
\maketitle{}

\section{Introduction}

Gamma-ray bursts (GRBs) can be detected up to the onset of reionization \citep[e.g.][]{tanvir, salvaterra} due to their brightness in the first few hours after the explosion \citep{lamb}. GRBs are transient sources followed by long lasting afterglows, emitting energy intensely across the full range of the electromagnetic spectrum. GRB afterglows are effective and informative probes for the detailed study of cosmic dust at high redshifts due to their simple power-law spectra, high luminosities and locations in star-forming environments. 

Interstellar dust has a crucial significance in the appearance and evolution of star formation in the early Universe. It is still under debate whether interstellar dust properties have evolved as a function of redshift. At high redshift ($z$ $>$$5$--$6$) it has been suggested that dust might originate in sources other than the evolved asymptotic giant branch stars that are the dominant source of dust production in the local Universe \citep{gehrz, dwek}. Previous studies reported that dust in cosmological objects at  $z > 6$ is predominantly produced in the ejecta of core-collapse supernovae (SNe), rather than the evolved stars which are missing on short timescales \citep{todini, morgan, nozawa, dwek, marchenko, hirashita}. Recently, however, \citet{valiante} calculated that the most massive asymptotic giant branch (AGB) stars could produce dust on time scales short enough to dominate dust production by $z\sim6$. Observationally, \citet{maiolino} found an unusual extinction curve in the most distant known broad absorption line quasar (BAL QSO) at redshift $z = 6.2$, consistent with what could be expected from dust produced in core-collapse SNe.

The progenitors of long-duration GRBs are known to be short-lived massive stars \citep{galama, hjorth, stanek, malesani}. GRB\,050904 at $z =6.295$ was a long duration GRB. It was extremely luminous and is the third most distant known GRB to date. GRB\,050904 was detected by \emph{Swift} on 2005 September 4 at $t_0=$ 01:51:44 UT \citep{cummings}. Substantial multi-wavelength follow-up was carried out simultaneously for GRB\,050904 with several ground based facilities. Previous analysis of the rest-frame UV afterglow found no evidence of dust \citep{tagliaferri, haislip, kann, gou}. Later \citet{stratta} re-examined the afterglow SED at different epochs and claimed the detection of dust absorptiontion with an extinction curve consistent with that used to explain the spectrum of the highest redshift BAL QSO, but inconsistent with the dust typical of the local Universe.

The claim of detection of SN-origin dust in GRB\,050904 is of fundamental importance to the question of the origin of dust in the early Universe, a very vexed problem for high redshift sub-mm galaxies (see, e.g. the discussion in \citealt{michalowski}). It was the first only direct observational evidence of dust from SNe in a high redshift star-forming environment. In this paper we review the relevant data to test whether dust is required by these observations and if so, what kind of dust is needed. The outline of the paper is as follows: In $\S$2 we describe the detailed multi-band spectral analysis of the afterglow of GRB\,050904 at different epochs. In $\S$3 we present results from the SED of the afterglow. In $\S$4 we discuss possible scenarios. In $\S$5 we provide our conclusions.

\section{Multi-wavelength observations of the afterglow}
\subsection{X-ray analysis}

\emph{Swift}'s X-ray Telescope (XRT) localized the afterglow of GRB\,050904. The XRT data (in the 0.3--10.0 keV energy range) were extracted and reduced using the HEAsoft software (version 6.4). We computed the X-ray spectra at three epochs, specifically 0.47, 1.25 and 3.4 days post-burst, chosen as the epochs with the best spectroscopic and photometric optical/near-infrared coverage. X-ray spectra at three epochs were created in a standard way using the most recent calibration files.

For our analysis, it is important to obtain an accurate estimate of the absolute flux for these X-ray spectra. The X-ray lightcurve is extremely variable, exhibiting long lasting flaring activity \citep{watson, cusumano, gendre}. The flares suggest two separate components, which may be due to a number of causes, possibly activity of the GRB central engine \citep[e.g.][]{burrows}. At late times the X-ray count rate is very low (see Fig.~\ref{FigFit}), therefore, we aim to get an accurate estimation of the flux which includes the uncertainty due to the flares. We fitted the afterglow lightcurve by assuming a smoothly broken power-law \citep{beuermann} to get an approximate overall X-ray lightcurve. The fit to the afterglow lightcurve results in at most a 30 -- 40$\%$ scatter around this fit at all epochs. We then normalized the X-ray spectra to the lightcurve fit at the relevant SED epoch. The procedure results in X-ray spectra with the best estimate of the slope and flux at the relevant SED epoch, and with an additional overall 30 -- 40$\%$ uncertainty on their absolute flux levels.

\subsection{Near-IR and optical imaging}

Several telescopes obtained photometric observations of the afterglow in the optical and near-infrared bands \citep{tagliaferri,haislip,price}. For a comprehensive investigation of the SED, we gathered optical and near-infrared photometry at three epochs. \citet{stratta} suggested unusual dust particularly on the basis of the $Z$-band observation at $\sim 0.5$ days, taken by the UK Infrared Telescope (UKIRT) equipped with WFCAM. Therefore, we re-reduced these $Z$-band data obtained at $t_0+0.47$ days. The object has low signal to noise. Our best estimate of the magnitude comes from point-spread function (PSF)-matched photometry, carried out using the DAOphot package within IRAF. For consistency we also carried out aperture photometry (with Gaia and IRAF/PHOT), and found the afterglow $Z$-band magnitude to be strongly sensitive to the chosen sky extraction annulus, being in some cases brighter by $\sim \mbox{half}$ a magnitude, although with large errors. This uncertainty is worth noting.

At $1.25$ days $z$-band photometry was obtained with the 8.2m ESO Very Large Telescope (VLT) by using the FOcal Reducer and low dispersion Spectrograph 2 (FORS2) $z$-Gunn filter. The object is detected with high signal to noise and using aperture photometry we could significantly reduce the error reported by \citet{tagliaferri}, which was likely mostly due to calibration issues.

At $3.3$ days after the burst, $z^\prime$-band photometry was carried out with the 8.2m Subaru telescope \citep{iye} using the Faint Object Camera and Spectrograph \citep[FOCAS;][]{kashikawa}. We re-analysed the photometric data using aperture photometry. We compared our FOCAS $z^\prime$ result with the Gemini South GMOS $z^\prime$ value at 3.16 days obtained by \citet{haislip}. Correcting for the time difference between the two observations using a broken power-law with $\alpha_1=0.72^{+0.15}_{-0.20}$, $\alpha_2=2.4\pm0.4$, $t_{\rm b}=2.6\pm1.0$ days \citep{tagliaferri}, we found these photometric measurements to be consistent within the statistical uncertainties.

Other photometric data were taken from the literature \citep{haislip, tagliaferri} when available at a time close to our nominal SED epochs ($J$-band at 0.49 days). All $H$ and $K_{\rm{s}}$ data as well as $J$ at 1.25 and 3.4 days were derived from the best-fit lightcurves of \citet{haislip} and \citet{tagliaferri}. At 1.25 days we used the $Y$-band photometry from the lightcurve computed in this band \citep{haislip}.

We corrected the observed magnitudes for extinction in the Milky Way (MW) using the \citet{schlegel} maps with $R_V=3.1$ and $E(B-V)=0.081$\,mag along the line of sight of the burst. Potential systematic uncertainties in the Galactic extinction correction have no significant effect on our results. The independent analysis of \citet{dutra} confirms the \citet{schlegel} maps for low $E(B-V)$. For the $z$-bands considered here, even assuming a 30\% uncertainty in the Galactic value would correspond to an extinction uncertainty of 0.03\,mag which is smaller than the statistical uncertainties we find for the extinction in the host galaxy (see Table~\ref{table:1}).  Effects in the $J$, $H$, and $K_{\rm{s}}$ bands will be smaller still. An under-estimate of the Galactic extinction would lead to a smaller host galaxy extinction than we find in the current analysis.

The $z$-Gunn, FOCAS $z^\prime$, GMOS $z^\prime$ and Sloan Digital Sky Survey (SDSS\footnote{\url{http://www.sdss.org/}}; \citealt{fukugita}) $z$ filters have almost the same profile across the whole band and extend much redder than the UKIRT $Z$ filter. It should be noted that since the Ly$\alpha$ absorption occurs in these bands, the filter wavelength coverage affects the observed magnitude significantly (see \S\,3.2).  Unless explicitly mentioned, in the rest of the paper we use the term ``$z$-band" to denote all of the three $z$ filters, i.e. UKIRT $Z$, FORS2 $z$-Gunn and FOCAS $z^\prime$.

\subsection{Grism spectroscopy}

The optical spectrum of the afterglow was obtained with FOCAS at the Subaru telescope and the spectroscopic data were retrieved from the SMOKA science archive facility\footnote{\url{http://smoka.nao.ac.jp/}} \citep{baba}. The afterglow was observed on 7 September 2005 for 4.0 hr. The exposure mid-time was 12:05 UT, corresponding to 3.4 days after the burst trigger \citep{kawai, totani}. The spectra were flux calibrated using the spectrophotometric standard star BD+28D4211 obtained on the same night. Individual spectra were combined following standard data reduction techniques using IRAF. The spectrum exhibits no flux below $\sim8900\,\AA$, consistent with a break due to Ly$\alpha$ absorption at redshift $z\sim6.3$ and the Ly$\alpha$ forest. The spectrum shows a flat continuum at the red wavelength end, revealing a series of metal absorption lines arising from different atomic species at $z=6.295$, and an intervening $\ion{C}{iv}$ system at $z=4.84$ \citep{kawai}. The observed spectrum was corrected for Galactic extinction by assuming the \citet{cardelli} extinction curve and as explained in \S\,2.2 above.

We implemented Voigt profile fitting to the 3.4 day Subaru spectrum using the FITLYMAN package in MIDAS \citep{fontana}. We measure a hydrogen column density of $\log{N_{\ion{H}{i}}}$ (cm$^{-2}$) $=21.62\pm0.02$, consistent with the value reported by \citet{totani}. It should be noted that $z^\prime$-band photometry and spectroscopy of the afterglow were obtained with FOCAS at 3.3 and 3.4 days, respectively.

\section{SED analysis}

\citet{stratta} studied the optical-UV rest-frame SED of the afterglow of GRB\,050904 at 0.5, 1 and 3 day epochs and found a deficit in the $z$-band at 0.5 and 1 days, and (less significantly) at 3 days, compared to the $JHK$ power-law extrapolation, claiming that dust reddening could explain the flux deficit. This required a SN-type extinction curve.

\subsection{Afterglow compound SED}

\begin{figure}
  \centering
   {\includegraphics[width=\columnwidth,clip=]{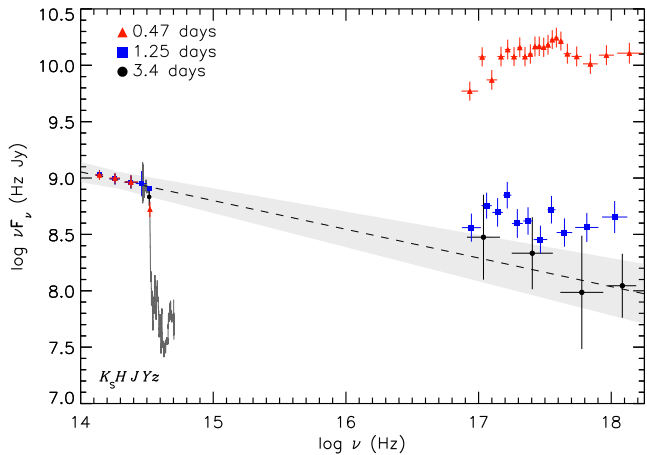}}
     \caption{Near-infrared to X-ray spectral energy distribution of the afterglow of GRB\,050904 at 0.47 (red triangles), 1.25 (blue squares) and 3.4 days (black circles) after the burst. The SED at 0.47 and 1.25 days is scaled to the $H$-band at 3.4 days. The solid grey curve represents the median-filtered optical spectrum at $t_0+3.4$ days. The black dashed line corresponds to a power-law fit to the near-infrared to the X-ray data at 3.4 days with a spectral index $\beta=1.25\pm0.05$. The grey shaded area represents the 1$\sigma$ uncertainty on the power-law.}
        \label{FigFit}
   \end{figure}

Knowledge of the SED can address the $z$-band flux suppression issue, therefore, we computed the near-infrared to X-ray SED of GRB\,050904 at three epochs, i.e. 0.47, 1.25 and 3.4 days. To facilitate comparison of the $z$-band flux, the SED at all epochs was normalized to the $H$-band flux, using the smoothly broken power-law presented by \citet{tagliaferri}. The normalized near-infrared photometry is generally consistent, but the X-ray spectra are much brighter at 0.47 and 1.25 days due to the intense afterglow flaring activity at these times. The consistency of the X-ray flux with the NIR SED extrapolation suggests that the X-ray afterglow at 3.4 days was relatively unaffected by flares. The composite SED of the afterglow of GRB\,050904 at three different epochs is shown in Fig.~\ref{FigFit}.

\begin{figure}
   \centering
   {\includegraphics[width=\columnwidth,clip=]{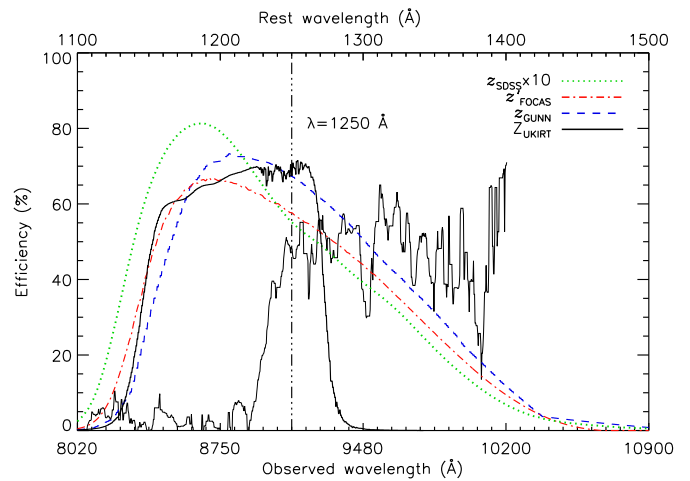}}
      \caption{Filter transmission curves of SDSS $z$, FOCAS $z^\prime$, FORS2 $z$-Gunn and UKIRT $Z$. The grey curve corresponds to the median-filtered optical spectrum at 3.4 days (arbitrarily normalized). The vertical triple black dot-dashed line represents the wavelength ($1250\,\AA$) where we computed the flux density (see \S\,3.2).}
         \label{Figmodel}
   \end{figure}

\subsection{Comparing the $z$-band filter responses}

Since the $z$-band photometry is strongly affected by the Ly$\alpha$ absorption, we performed spectro-photometric analysis by utilizing the total effective filter transmission functions including detector responses (Fig.~\ref{Figmodel}). We use the following method that allows for a clean comparison of the different $z$-band magnitudes of the afterglow, taken at 0.47, 1.25, and 3.4 days after the burst, with the filters UKIRT $Z$, VLT $z$-Gunn, and Subaru $z^\prime$, respectively. The method essentially constructs the SEDs of stars in the field and uses these to make a direct comparison of the afterglow magnitudes at each epoch.

First, in each afterglow image we select several non-saturated reference stars with known SDSS and Two Micron All Sky Survey (2MASS\footnote{\url{http://www.ipac.caltech.edu/2mass/}}; \citealt{skrutskie}) magnitudes. Using the 2MASS $J$-band, and the SDSS $z$ and $i$, we construct a rudimentary SED for each reference star, where we convert the magnitudes to flux densities (in $\rm{erg}$ $\rm{s}^{-1}$ $\rm{cm}^{-2}$ $\AA^{-1}$) at the central wavelength of the SDSS and 2MASS filters. We connect these flux densities with a linear interpolation, and integrate the reference star SED convolved with the filter response curve relevant to that image, retrieving the band-integrated flux in $\rm{erg}$ $\rm{s}^{-1}$ $\rm{cm}^{-2}$. Second, from the image we measure the counts for the reference stars using aperture or PSF photometry, determining the conversion factor between counts and flux. Using this factor, we eventually compute the (band-integrated) afterglow flux from its measured counts. We used several comparison stars to evaluate the accuracy of the procedure. At 0.47, 1.25, and 3.3 days, we find a scatter of 0.02, 0.04, and 0.02\,mag using 8, 10, and 5 reference stars, respectively. The small scatter confirms the robustness of our method.

In the deep FORS2 and FOCAS $z$-band images, the brightest stars are heavily saturated, and suitable reference stars are lacking since the fainter stars have large uncertainties in the SDSS and 2MASS catalogues. Therefore, we calibrated a set of faint stars using the UKIRT $J$ and $z$-band images, based on the 2MASS and SDSS catalogs. For the $z$ band, due to the difference in the UKIRT and SDSS filters, appropriate color terms were taken into account, achieving a photometric accuracy of $\approx 0.02$\,mag. Given the higher sensitivity of the SDSS in the $i$ band, suitable calibrators for the VLT and Subaru images were available directly from the SDSS catalog. Note that our calibration is entirely based on the SDSS and 2MASS catalogs, therefore, our analysis is not dependent on the sky conditions when the data have been acquired.

The third and final step is to convolve the relative afterglow spectral shape (as measured from the Subaru spectrum that was obtained at 3.4 days) with the three different $z$-band filter response curves, where the spectrum is rescaled in absolute terms to recover the band-integrated flux (in $\rm{erg}$ $\rm{s}^{-1}$ $\rm{cm}^{-2}$) determined for each epoch (see above). We note that this method does not rely on the absolute flux calibration of the Subaru spectrum; it merely uses the photometry to rescale it, and therefore the errors only include the errors in the aperture/PSF photometry, the error from the conversion factor,  and the Subaru noise error when convolving it with the filter response curves. After rescaling of the spectrum, the afterglow flux density (in $\rm{erg}$ $\rm{s}^{-1}$ $\rm{cm}^{-2}$ $\AA^{-1}$) can be compared at any pivot wavelength, after the $H$-band normalization. At all epochs we computed the flux density at $\lambda_{\rm{rest}}=1250\,\AA$, which was selected since it is close to the peak of all the involved filter transmission curves (see Fig.~\ref{Figmodel}), and is separated from the metal absorption lines visible in the spectrum \citep{kawai}.

The single but important assumption in this method is that the spectral shape of the afterglow is not changing from 0.47 to 3.4 days. This is in some sense the null hypothesis that we are trying to test: dust destruction would produce a change in the relative spectral shape, and would therefore produce a change in the $z$-band brightness relative to the $H$-band normalization. Other effects, such as a variable $\ion{H}{i}$ column, or a change in the spectral slope due to e.g. the cooling frequency crossing the $z$-band, could also in principle cause such a change. However, if the resulting afterglow $z$-band brightness (normalized to the $H$-band) between 0.47 and 3.4 days does not show a significant change, then this would provide strong support for the null hypothesis that the spectral shape is not changing.

Following the above procedure, we find the afterglow to have a flux density at $1250\,\AA$ of $14.7\pm1.32$, $9.41\pm0.24$ and $2.13\pm0.22$ $\mu$Jy at 0.47, 1.25 and 3.3 days, respectively. We find that the normalized 0.47 day UKIRT $Z$-band brightness is $0.27 \pm 0.18$\,mag fainter than the FOCAS $z^\prime$-band brightness at 3.3 days. At 1.25 days, the normalized FORS2 $z$-Gunn brightness is brighter by $0.17 \pm 0.15$\,mag compared to the FOCAS $z^\prime$-band brightness at 3.3 days. The uncertainties here also include the uncertainties in the normalization, i.e. the errors in the $H$-band photometry (0.06\,mag). Therefore, there is no evidence for variability of the spectral shape around the $z$ band. In particular, after taking into account the appropriate filter shapes and color effects, there is no significant deficiency of flux in the $z$-band flux at 0.47 days compared to later epochs \citep{haislip,stratta}.

\section{Discussion}

At 0.47 days post-burst, we find a flux deficit in the UKIRT $Z$-band compared to the 3.3 days Subaru photometry that is only significant at  $<1.5\sigma$ level. This low significance result, combined with the difficulty in determining the $Z$-band magnitude at 0.47 days alluded to in \S 2.2, suggests that a change in the spectrum between 0.47 and 3.4 days does not have strong observational support. If the effect were real then such a flux deficit could be explained by: (i) dust extinction as suggested by \citet{stratta} with a SN-origin extinction curve, or (ii) gas absorption. Previously \citet{haislip} also suggested that absorption due to molecular hydrogen could give rise to the $Z$-band flux deficit at 0.47 days.

\subsection{Dust in the GRB\,050904 host galaxy}

The claim of SN-type dust in GRB\,050904 is important because of the possibility of observing the evolution of cosmic dust at high redshift. \citet{stratta} suggested SN-type dust extinction in the host galaxy of GRB\,050904 with an extinction curve inferred for a BAL QSO at $z=6.2$ \citep{maiolino}. The unusual extinction curve is rather flat at longer wavelengths and steeply rises at $\lambda<1700\,\AA$. The best-fit estimates of \citet{stratta} of the extinction at 3000\,\AA\ in the rest-frame, $A(3000\,\AA)$, were $0.89\pm0.16$, $1.33\pm0.29$, and $0.46\pm0.28$\,mag at 0.5, 1, and 3 days, respectively. 

\begin{figure}
  \centering
   {\includegraphics[width=\columnwidth,clip=]{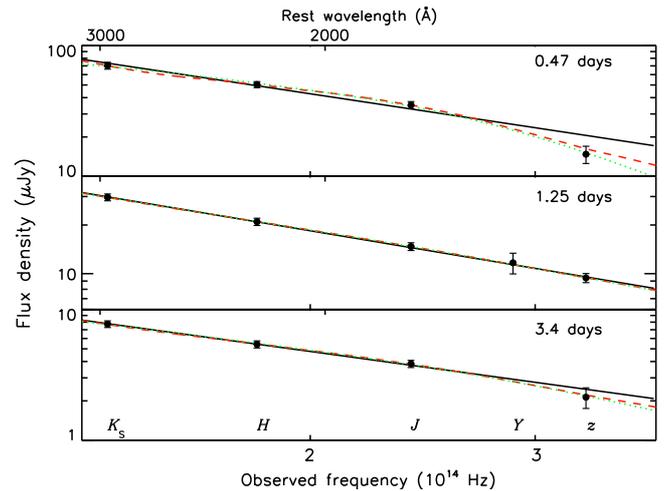}}
     \caption{Near-infrared spectral energy distribution of the afterglow of GRB\,050904 at 0.47 ($top$ $panel$), 1.25 ($middle$ $panel$) and 3.4 days ($bottom$ $panel$) after the burst. The observed data are corrected for Galactic extinction (\S\ 2.2). The corresponding bands are identified in the bottom panel. The solid, dashed, and dotted lines represent the best fit with a power-law, a power-law with SN dust, and a power-law with SMC dust, respectively. At 1.25 days, the three lines almost overlap.}
        \label{Figex}
   \end{figure}

It is clear from our broad-band SED at 3.4 days (see Fig.~\ref{FigFit}) that the extrapolation of the near-infrared power-law is consistent with a single power-law to the X-ray spectrum, i.e.\ consistent with both the slope and flux level of the X-ray spectrum at that time. We can also clearly see that there is no evidence in the flux-calibrated optical/near-infrared spectrum at 3.4 days for any extinction -- the continuum just redward of the Ly$\alpha$ absorption is consistent with the single $JHK$ power-law. Both facts mean that there is no evidence for dust extinction at 3.4 days. We fitted a dust-attenuated power-law using a dust model for the Small Magellanic Cloud \citep[SMC, $R_V=2.93$;][]{pei} and the SN-origin extinction model of \citet{maiolino} to the 0.47, 1.25 and 3.4 day $zYJHK_{\rm{s}}$ data (from the $z$-band, we compute the flux density at $\lambda_{\rm{rest}}=1250$\,\AA). The best fit parameters are reported in Table~\ref{table:1}. With our revised $z$-band photometry, extinction at the level suggested by \citet{stratta} can be ruled out at all three epochs (see Fig.~\ref{Figex}). In no case the computed absorption is significant at more than 1.5$\sigma$ level.

\begin{table}
\begin{minipage}[t]{\columnwidth}
\caption{Best fit parameters of the SED at different epochs.}      
\label{table:1} 
\centering
\renewcommand{\footnoterule}{}  
\begin{tabular}{@{}c c c c c@{}}   
\hline\hline                        
Days & Model & $\beta$ & $A(3000\,\AA)$ & $A_V$ \footnote{The SN-origin extinction curve has been only computed in the range $\lambda_{\rm{rest}}=1000-4000\,\AA$, hence it is not possible to provide $A_V$.}\\
 &  & & (mag) & (mag)\\ 
\hline
0.47 & PL & $1.28 \pm 0.11$ & \mbox{\ldots} &  \mbox{\ldots} \\
		& PL+SN & $1.22 \pm 0.24$ & $0.3\pm0.22$ & \mbox{\ldots} \\
		& PL+SMC & $1.23 \pm 0.08$ &  $0.1 \pm 0.07$ & $0.05 \pm 0.04$ \\ [5pt]
\hline
1.25 & PL & $1.24 \pm 0.09$ &  \mbox{\ldots} &  \mbox{\ldots} \\
		& PL+SN & $1.27 \pm 0.2$ & $0.05\pm 0.11$ &  \mbox{\ldots} \\
		& PL+SMC & $1.17 \pm 0.51$ & $0.01 \pm 0.04$ & $0.01 \pm 0.02$ \\ [5pt]
\hline
3.4 & PL & $1.25 \pm 0.05$ &  \mbox{\ldots} &  \mbox{\ldots} \\
		& PL+SN & $1.23 \pm 0.21$ & $0.22 \pm 0.24$ & \mbox{\ldots} \\
		& PL+SMC & $1.24 \pm 0.07$ & $0.056 \pm 0.059$ & $0.042 \pm 0.044$ \\ [5pt]
\hline
\end{tabular}
\end{minipage}
\end{table}

Extinction-correcting the 1.25 day SED at the level fitted by \citet{stratta} makes its extrapolation overshoot the X-ray spectrum, hinting that $\sim1$\,mag of extinction at 3000\,\AA\ is not required. More importantly, the $Y$-band photometry, with a central wavelength of 1400\,\AA\ in the rest-frame, at 1.25 days \citep{haislip}, is consistent with the near-infrared power-law extrapolation. Such consistency would not be expected in the \citet{stratta} dust hypothesis since $A(1400\,\AA)$ is about 1.75 times the $A(3000\,\AA)$ in the \citet{maiolino} model, and the $Y$-band photometry should therefore lie a factor of 2 below a power-law extrapolation, while it does not (Fig.~\ref{Figex}), though its error is large. As it can be seen in the middle panel of Fig.~\ref{Figex}, the SED at 1 day follows a simple power-law and provides strong constraints on dust absorption. Again, it seems likely that not only is there no evidence for SN-type extinction in GRB\,050904 after 1.25 days, but that there is no evidence for any dust extinction at all at $\sim1$ day or later.

There are also strong arguments against a SN-origin dust interpretation at 0.47 days. While dust reddening has been unequivocally observed in lower redshift GRBs (e.g.\ \citealt{kann06};  \citealt{fynbo09}; GRB\,050401: \citealt{watson06}; GRB\,991216: \citealt{vreeswijk06}; GRB\,050408: \citealt{foley}; \citealt{ugarte}; GRB\,070802: \citealt{eliasdottir}; GRB\,080607: \citealt{prochaska09}), so far SN-origin dust has never been seen before in any GRB host. Moreover there is no compelling evidence of dust extinction in any GRB beyond $z=5$. A possible exception is GRB\,071025 \citep[which has a photometric redshift $4.4 < z < 5.2$][]{perley}, which shows indications of a significant dust column. Notable are the two bursts at higher redshift than GRB\,050904, i.e. GRB\,080913 at $z=6.7$ \citep{greiner}, and GRB\,090423 at $z=8.2$ \citep{tanvir, salvaterra}, neither of which show any sign of extinction. Second, given that dust can be excluded at $t>1$ day, having non-zero absorption at $t = 0.47$ days would require time-varying dust extinction, which has never been observed in any burst. If due to dust destruction, we would expect reddening variations to be associated with intense episodes of emission, while there is no optical flaring or any significant feature in the restframe-UV lightcurve in this period that could be responsible for such dust destruction \citep[see][]{haislip,tagliaferri}, and most dust destruction scenarios sublimate dust on timescales of only a few minutes after the burst at most \citep{perna03, fruchter}. \citet{stratta} suggested that varying extinction may also indicate that the emitting region had become larger than the obscuring cloud. While this cannot be excluded, such a geometry requires some tuning of the cloud and fireball parameters. The claim of dust in the host galaxy of GRB\,050904, with an unusual extinction curve, relying principally on a smaller (0.3\,mag) and $<2\sigma$ flux deficit in a photometric observation, is not the most likely explanation. The most likely hypothesis is simply systematic uncertainties related to the $Z$-band calibration.


It is worth noting however that time-variable dust with an unusual extinction curve is not even the simplest explanation even if the original analysis had been reliable. Given that we know from the optical spectrum that a large quantity of gas is present in the system, a variability in the gas column density at early times is a less tortured hypothesis.

 \subsection{X-ray absorption}

The X-ray spectral analysis suggests a high metal column density in the afterglow of GRB\,050904 \citep{watson, campana}. Time-resolved X-ray spectroscopy reveals that the column density of metals within the first few hours is highly variable \citep{campana, cusumano, gendre}. Due to the rapid changes in the X-ray spectrum this apparently variable column may be an artifact of the changing intrinsic spectrum resulting in a downturn at soft energies that disappears at later times \citep[see][]{butler}. However, even if the change in the soft X-rays is really due to a variable column density, i.e.\ due to increasing ionization of the metals, this effect occurs at early times ($\lesssim 10^3$\,s) and cannot support the idea of dust destruction after 0.5 days. Indeed, a varying metal column density at $<1000$\,s argues against dust destruction at 0.5 days. If the varying metal column density is a real effect, destruction of any dust associated with the high metal column should have been completed long before 0.5 days. As a more general point, the optical and X-ray fluxes are at least one to two orders of magnitude lower after 0.5 days than before 1000\,s. It is difficult to construct a scenario in which significant dust destruction occurs in the interval 0.5--3 days that did not occur before in the absence of a huge flare in the UV--X-ray, something which is not observed.

\subsection{Gas-to-dust ratio}

GRBs typically occur in host galaxies with high gas to dust ratios (e.g. \citealt{jensen, galama01, hjorth03, stratta04, eliasdottir}). The \ion{H}{i} column density of the host of GRB\,050904 is very large while $A_V$ is small. Using our limit on (SMC-type) dust at 1.25 days, $A_V\lesssim0.05$\,mag at $95\%$ confidence, leads to a high gas-to-dust ratio $N(\ion{H}{i})/A_V \gtrsim 8.3\times10^{22}$\,cm$^{-2}$\,mag$^{-1}$. The Galactic relation between \ion{H}{i} column density and dust reddening is $N(\ion{H}{i})/A_V=4.93/R_V \times10^{21}$\,cm$^{-2}$\,mag$^{-1}$ \citep{diplas}. Correcting for metallicity at 3.4 days ($Z=0.1Z_\odot$; \citealt{kawai}), this implies an $N(\ion{H}{i})/A_V$ ratio limit 5 times the Galactic value. A comparison with the SMC \citep{gordon}, which has a similar metallicity to the environment of GRB\,050904, yields  a gas-to-dust ratio which is also more than 5--10 times larger in the host of GRB\,050904.

\subsection{The origin of dust in the early Universe}

In the local Universe, the major sources of interstellar dust are AGB stars, the lower mass ranges of which require $\gtrsim 1$\,Gyr to evolve to produce dust \citep{dwek}. It has been suggested that for sources with large dust masses such as sub-mm galaxies, due to the short time available at $z\gtrsim5$, an alternative source of dust is required and that core-collapse SNe could dominate dust formation at these times \citep{todini, morgan, nozawa, dwek, marchenko, hirashita}. However, more complete theoretical models including dust destruction by supernova  shock or grain growth/destruction in the interstellar medium obtain yields that are $\lesssim0.01$\,$M_\odot$ per SN \citep{bianchi}, consistent with almost all observations of nearby SNe \citep{wooden, elmhamdi, meikle, blair, sakon}. This is too little to produce the quantities of dust observed at high redshift. Recently \citet{valiante} argued that on short timescales massive AGB stars could form much of the dust, depending on the assumed initial stellar metallicity and star formation history. The galaxy-SED modelling of sub-mm--selected galaxies of \citet{michalowski} suggests dust-formation timescales of order tens of millions of years in a few cases at $z\gtrsim4$, which would clearly preclude even high-mass AGB dust-formation. While intriguing, these cases may be affected by active galactic nuclei (AGN) contamination and must be treated cautiously.

Observationally, after our analysis here of the afterglow of GRB\,050904, the detection of a peculiar extinction curve in a BAL QSO spectrum at $z = 6.2$ \citep{maiolino} remains the only direct evidence for dominant SN-origin dust in the early Universe (but see recent work by \citet{perley}). While the observational analysis of \citet{maiolino} is carefully done, the relatively narrow wavelength coverage, the presence of strong, broad absorption and emission lines that dominate over the continuum at the blue end of the spectrum, and the use of composite QSO spectra, leave the result awaiting further confirmation. Furthermore, it is difficult to exclude that the dust is affected by the central AGN itself \citep{perna03}, so that the extinction curve may not tell us a lot about the origin of that dust.

\section{Conclusions}
In this work we reinvestigated the afterglow of GRB\,050904 at 0.47, 1.25 and 3.4 day epochs to understand stellar environments and interstellar dust at high redshift. We find that the afterglow SED can be reproduced at all epochs without any dust extinction. The previous finding of dust extinction requiring a SN-type extinction curve by \citet{stratta} relies mostly on a $Z$-band photometric point at 0.47 days which we find has calibration difficulties and with our new accurate analysis technique we find the flux deficit to be both smaller and less significant than reported by previous studies. We can reasonably exclude the presence of substantial quantities of any type of dust in this GRB host galaxy at all epochs. We therefore conclude that there is no significant evidence of dust extinction in the afterglow of GRB050904.

\begin{acknowledgements}

The Dark Cosmology Centre is funded by the Danish National Research Foundation. Based in part on data collected at  Subaru Telescope and obtained from the SMOKA archive, which is operated by the Astronomy Data Center, National Astronomical Observatory of Japan. Our special thanks to Giorgos Leloudas for helpful discussions. We are grateful to Tomonori Totani, Kentaro Aoki and Takashi Hattori for helping us in the Subaru data re-reduction. The authors thank the referee for very positive and constructive comments.



\end{acknowledgements}

Note added post-submission: A recent paper by \citet{perley} reports significant SN-origin dust extinction in GRB\,071025 at $z$ $\sim$ $5$ \citep{perley}. We note that \citet{perley} also independently attempted to model the dust profile of GRB\,050904 and found that the data are consistent with no extinction at all.
\bibliographystyle{aa}
\bibliography{GRB050904.bib}

\end{document}